\documentclass[runningheads]{svmult}

\usepackage{makeidx}   
\usepackage{graphicx}  
\usepackage{subeqnar}  
\usepackage{multicol}  
\usepackage{physprbb}  
\makeindex             

\begin{document}
\title*{Outflows in Regions of Star Formation}
%
%
%
%
\titlerunning{Outflows from YSOs}
%
\author{Ren\'e Liseau\inst{}
 }
\authorrunning{Ren\'e Liseau}
%
%
\institute{Stockholm Observatory, AlbaNova University Center, SE-106 91 Stockholm, Sweden 
e-mail: rene@astro.su.se
 }

\maketitle              

\begin{abstract}
The high spatial and spectral resolution offered by the new generation of infrared
spectrometers at ESO is optimally suited for the observational study of outflows from young 
stellar objects. Models of interstellar shock waves would benefit from observations of spectrally
resolved line profiles. This applies also to attempts of measuring the rotation
rates of jets very close to their driving source, which in general suffer considerable 
extinction. Observations of forbidden lines of ionised iron, [Fe\,II], could be used to accomplish this.
The possibility of using rotational lines of molecular hydrogen, H$_2$, to
study the temporal evolution of outflow and disk gas is discussed. Similarly, high resolution 
IR observations of fluorescent water lines, H$_2$O, open up the possibility to access outflow and disk
water from the ground.
\end{abstract}

\section{Introduction}

Rapid stellar evolution is generally accompanied by intense mass 
loss phenomena. During both the pre-main-sequence (Richer et al. 2000) 
and the post-main-sequence evolution (Sahai 2002),
there are phases during which the mass loss has a distinctly non-spherically symmetric 
distribution.
The overall geometry is often best described as hourglass-shaped or wide-bipolar,
although in several cases the flows can take the form of narrow jets. 

Actually, the observation of bipolar jets is not limited to stellar outflows 
but these are also commonly observed in active galactic nuclei (AGN). 
In fact, pictorial models of protostars and AGN are indistiguishable, where
the salient features include a central object, surrounded by an active accretion 
disk and a perpendicularly directed bipolar jet (Blandford \& Payne 1982). These
morphological similarities could suggest the same physics (momentum re-distribution) 
to operate, over a wide range
of physical scales, and could provide a connection between stellar and
extragalactic astrophysics. This familarity is further illustrated by the fact 
that Chandra and XMM-Newton recently have detected the flows from pre-main-sequence 
objects in (likely thermal) X-rays
(Pravdo et al. 2001, Favata et al. 2002). This carries the study of these flow phenomena also into the 
traditionally home domain of AGN, i.e. that of high energy astrophysics.

Although many of the details remain uncertain, there is little doubt that the
energy is ultimately extracted from the gravitational and magnetic fields of the 
source. The understanding of outflow phenomena is thus intimately 
related to the understanding of the overall energy and momentum budget of the star/galaxy 
formation process and implies, as such, more general ramifications.

Extragalactic and post-main-sequence stellar outflows, as well as
circumstellar disks and high-mass star forming regions, are addressed elsewhere in these proceedings.
Here, we shall focus on outflows from low-mass objects in their early stages of pre-main-sequence evolution.

\section{Outflows in Regions of Low-Mass Star Formation}

\subsection{Angular Momentum: Stellar Rotation, Disks and Outflows}

Although stellar rotation is most likely controlled by local processes, 
as observationally indicated by the randomly oriented rotation axes,
even galactic differential rotation alone would suffice to result in break-up of
the contracting protostars, unless their angular momentum would be  
redistributed with high efficiency. Viscous disk accretion is commonly invoked to 
accomplish this, through disk locking or a stellar wind (Rebull, Wolff \& Strom 2004; 
Matt \& Pudritz 2004). However, the observational evidence appears only
partially supportive for this widely accepted scenario and a convincingly strong 
relationship between stellar rotation and the presence of protostellar
disks in young low-mass pre-main-sequence stars has yet to be established 
(Mathieu 2003).

The presence of accretion disks is generally also required as a pre-requisite
for theoretical models of outflows. Recently, models have been 
developed which simultaneoulsly address the physics of both gravitational collapse 
and the origin of outflows (see, e.g., Tomisaka 2002 and references therein). These
models take also rotation and magnetic fields explicitly into account, needed
for the angular momemtum transport in the viscous accretion disks (Balbus \& Hawley 1991, 
Stone et al. 2000) or off the disks (K\"onigl \& Pudritz 2000, Shu et al. 2000). 
According to Contopoulos \& Sauty (2001), the removal of angular momentum could in 
certain circumstances become extremely efficient, viz. expressable in terms of
the ratio of the mass loss to the mass accretion rate, 
{\it \.{M}}$_{\rm loss}$/{\it \.{M}}$_{\rm acc} \rightarrow 1$. 

\subsection{Observed Physical Characteristics of Outflows}

Recent reviews on this and related topics are by Reipurth \& Bally (2001)
and those collected in the volume of Protostars \& Planets IV, 
including those by Eisl\"offel et al. (2000),
Hartigan et al. (2000), K\"onigl \& Pudritz (2000), Richer et al. (2000) and by
Shu et al. (2000). 

To summarise shortly, a generalised theory of HH/jet and molecular flows has to account for 
at least the following observed properties: 
{\it length scales} of $\sim 10^{-4}$\,pc to a few parsec,
{\it opening angles} of ${\sim} 1^{\circ}$ to $> 90^{\circ}$,
{\it space velocities} of some km\,s$^{-1}$ to several $10^2$\,km\,s$^{-1}$,
{\it gas kinetic temperatures} of $\stackrel {<}{_{\sim}} 10$\,K to a ${\rm few} \times 10^6$\,K.
Further, based on the assumption of momentum conservation, derived mass loss rates scale with the radiative 
luminosity of the central source(s), i.e. 
{\it \.{M}}$_{\rm loss} \propto L_{\rm mech} \propto L_{\rm bol}^{\,\,p}$, 
where $p \sim 0.6$ to 1. In regions of low mass star formation,
observed (line of sight) magnetic field strengths are generally less than $30\,\mu$G on the arcminute 
scale (Crutcher et al. 1993; Crutcher \& Troland 2000).
It should be realised that only a few of the above parameters have been determined with 
sufficiently high accuracy. In addition, time dependent effects cannot be ignored as several flows
show evidence for time variablity.

\subsection{Origin of Outflows: Driving Mechanisms}

Observed outflows display a wide range of effective collimation and
a dichotomy of theoretical models for the driving of outflows has emerged, which 
differentiates between wide-angle winds (Shu et al. 1991, 2000) and collimated jets 
(Stahler 1994). In the former case, a stellar wind or a wind from the disk surface is 
transferring momentum to the surrounding molecular material, whereas in the latter case,
molecular outflows are thought to be molecular material entrained through
a turbulent layer of a stellar jet/Herbig-Haro flow. HH-flows are often observed to be spatially 
associated with molecular outflows. Observations with the ISO-LWS have provided convincing evidence that
HH-flows and molecular outflows have a common physical cause (Liseau et al. 1997; Saraceno et al. 1998).

All model families tend to reproduce observed outflow properties, such as
the mass-velocity relation, $m(v) \propto v^{-\gamma}$ with $\gamma \sim 1.8$, 
and the `Hubble Law', $v(r) \propto r^{\alpha}$ with $\alpha \sim 1$. Recent supporting
arguments for the wide-angle wind hypothesis have been provided by Matzner \& McKee (1999) and by
Gardiner, Frank \& Hartmann (2003). However, questions regarding the launching and collimation
are left open. Downes \& Cabrit (2003) argue in favour of the jet-driving hypothesis, but their
main results seem not very different from those of the wind-driven models by Lee et al. (2000). 

As discussed by Ray (2000; see also: Cabrit, Raga \& Gueth 1997), 
both model categories have their strengths - but also their weaknesses. 
The respective shortcomings might be overcome, if the possibility of temporal evolution of the outflows 
is considered, where initially highly collimated molecular `jet-flows' evolve into increasingly
less collimated outflows as they age. 

\section{Open Questions}

\subsection{Physics}

\subsubsection{Origin of Matter Outflow}
 
The, by far, most outstanding problem concerns the origin of the outflows. To gain further insight,
direct observation of the regions of acceleration would most likely be required. MHD models make detailed
predictions of where this should happen and what the angular velocity of the rotating flow should be
(Bacciotti et al. 2002; Presenti et al. 2004).
The relevant scale is set by the Alfv\'en radius, the physical size of which depends on the 
details of the favoured model. But somewhat losely, one may distinguish
X-winds, which are generated within only a few $R_{\odot}$ ($10^{-2}$\,AU) of the central source,
and disk winds, which are launched within a fraction of an AU to some AU. At the typical distance 
of outflows in low mass star formation regions of some $10^{2}$\,pc, acceleration regions are 
thus expected to subtend angles of less than a milliarcsec to a few tens of milliarcsec.
A related issue is that of the mechanism of initial flow collimation, occuring over similarly small 
spatial scales, whether MHD or not.

\subsubsection{Temporal Evolution}

Of great relevance to the physics of star formation is of course the evolutionary status of the central 
driving source, be it single be it binary/multiple, the latter seemingly more likely (M.\,Barsony, 
private communication). Age estimates of outflow engines range from only a few
thousand to about a million years. One of the key questions is to what extent
the properties of observed flows, showing great diversity, can be directly linked to the pre-main-sequence
age of their driving sources. In particular, one would wish to establish the value 
and the time dependence of the `$f$-factor' on an observationally firm basis. 
This factor, defined as {\it \.{M}}$_{\rm loss}$\,=\,$f \times$\,{\it \.{M}}$_{\rm acc}$, 
is a measure of the relative importance of mass inflow and mass outflow for an individual source. 
MHD models concur with a value of $f \sim 0.3$. In order to understand the underlying physics of the
outflows at their root, it is thus important to learn, whether time evolution alters this `constant'
in a systematic way.

\subsubsection{Environmental Effects}
 
An item of great importance regards the understanding of the stability of interstellar clouds against 
their self-gravity,
i.e. cloud lifetimes are many times longer than their free-fall time scales. Observed cloud `turbulence' 
is generally believed to accomplish this, although the energy source, injecting and  maintaining 
the turbulent motions, has yet to be uniquely identified. Outflows from young stellar objects represent
one of the known prime candidates. 

\subsubsection{Shock Physics}

Outflows also offer the opportunity to study the 
physics of a particular class of interstellar shock waves. These tend generally to be radiative
and reasonably thermally stable. Specific problems include upstream-downstream communication,
particle acceleration, shock cooling and grain destruction processes.

Specifically for the jet-driven outflow models, the problem exists how to account for the wide angles of
observed molecular outflows. These flows are more often the rule than the exception. In other words,
the mechanism for the momentum transfer in the transverse direction has yet to be uniquely identified.

In relation to this, it is important to provide firm observational data for the internal (jet) and external
(surrounding medium) densities, as the driving of the molecular flows could be accomplished only by
underdense jets. The development of instabilities in the molecular flow (high velocity knots or `bullets',
see Bachiller 1996), possibly related to this issue, is not well understood either.

\subsection{Chemistry}

\subsubsection{Environmental Effects}

As mentioned in the previuos paragraph, outflows affect their environment physically, but 
their molecular clouds may also transform chemically. On a global scale, the effects of outflows
on the chemical evolution of their host cloud have been addressed by Bergin, Melnick \& Neufeld (1998).
These models focus on the relative abundance, as a function of the time, of primarily a few key species 
(chemistry and cooling) such 
as O$_2$ and H$_2$O. At the time, these models appeared elegant and highly attractive. Unfortunately, and
unexpectedly, {\it Nature} did not agree, as observations with SWAS (Snell et al. 2000; Goldsmith et al. 2000) 
and Odin (Hjalmarson et al. 2003; Pagani et al. 2003) have demonstrated. Both molecules were found
to be severely underabundant with respect to the (steady state) chemical model predictions and several 
proposals to solve this problem have been offered, some of which have other flaws (see the references 
in Bergin et al. 2000 and in Pagani et al. 2003).
Furthermore, in view of the great variety of the proposed solutions, this issue seems far from settled.

\subsubsection{Shock Chemistry}

On the local scale of the outflows, the study of the chemistry behind the shock waves is of relevance. What kind of
shock (dissociative $J$ and/or non-dissociative $C$) will develop, depends on the flow itself and the medium it runs into. 
The predicted chemistry will be correspondingly different in these cases. The understanding of the chemistry is 
important per se, but also because of its feed-back into the physics (via the energy balance/cooling, which in turn
feeds back into the chemical reactions), thereby providing 
the proper spectral diagnostics for the observation of the flow (which lines, their intensity and shape). 

A problem of theoretical models is their susceptibility to the choice of the boundary conditions and input parameters. 
In addition, the shocked flow might not have had the time to establish chemical equilibrium, implying that time
dependence cannot be ignored. 

For instance,
observed abundance ratios $X$(OH)/$X$(H$_2$O) in a number of outflows can be used to illustrate this point.
Steady state $C$-shock models are inconsistent with the observations and their relative underproduction of OH
could be explained in two alternate ways. (1) The inconsistency arises as a consequence of a time dependence effect, 
i.e. the flows are too young to have converted sufficient amounts of O into H$_2$O (Richer et al. 2000), or (2) 
is due to an incorrectly chosen input parameter, i.e. the actual ionisation rate is significantly higher, perhaps
because of the presence of X-ray sources (Wardle 1999; Larsson, Liseau \& Men'shchikov 2002). 

The resolution of the OH- and similar problems will have to make use of detailed comparison of observational data with
the results of carefully selected, time dependent chemical model calculations.

\subsubsection{Temporal Evolution}

However, non-equilibrium systems are not only `bad', since their sensitivity can be turned to an advantage: if
one is convinced that one has picked the right model of the source, then this model can be used for timing 
purposes in order to identify various
events occuring during the evolution of the source. Hence, with regard to the chemical models, the a priori
problem lies in identifying the most plausible ones, so that they can also be used as chronometers.
The ortho-to-para state ratio of, e.g., H$_2$ is a chief example to achieve timing observationally in outflows.

\subsubsection{Chemical Complexity}

A very modest number of outflows shows spectra with a rich variety of molecular lines (L\,1157: Bachiller et al. 2001; 
BHR\,71: Bovill, Bourke \& Bergin 2002), whereas most observed outflows do not.
As the chemically enriched ouflows are more collimated than on the average and they emanate from sources believed
to be very young (Class\,0), it is reasonable to interpret the enrichmment as an age (or rather youth) effect.
Observed abundance gradients along the flow would be consistent with this idea, being due to the chemical evolution 
of the shocked gas (Bachiller et al. 2001). 

It is at present not clear, whether the known small fraction of Class\,0 associated outflows showing enhanced
chemical abundances is the result of an observational selection effect. Foregoing the results of ongoing and 
future observations, the question of outflow evolution needs to be addressed also from the chemical point of
view.

\section{High Resolution IR Observations}

Much of the gas in molecular outflows is at low temperatures ($T$ a few tens of Kelvin) and is most readily observed at 
submm/mm wavelegths. In this regime, the contrast between the outflow and the hosting molecular cloud is often low at
velocities close to the systemic velocity, thus confusing the observation of the outflow gas with that of the
host cloud. Higher excitation material, observable at shorter wavelengths (e.g., in the IR), circumvents this
difficulty.   

\begin{figure}
\begin{center}
\includegraphics[width=0.8\textwidth]{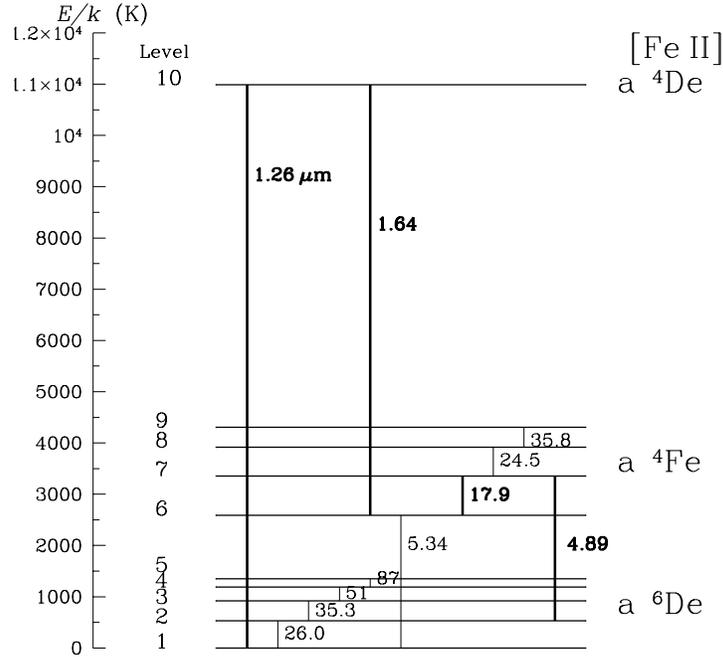}
\end{center}
\caption[]{Energy level diagram for the ten lowest levels of [Fe\,II], with the energies expressed in Kelvin.
Wavelengths in $\mu$m are indicated next to the shown transitions. The bold lines 
identify transitions discussed in the text  }
\label{feii_levels}
\end{figure}

\begin{figure}
\begin{center}
\includegraphics[width=0.8\textwidth]{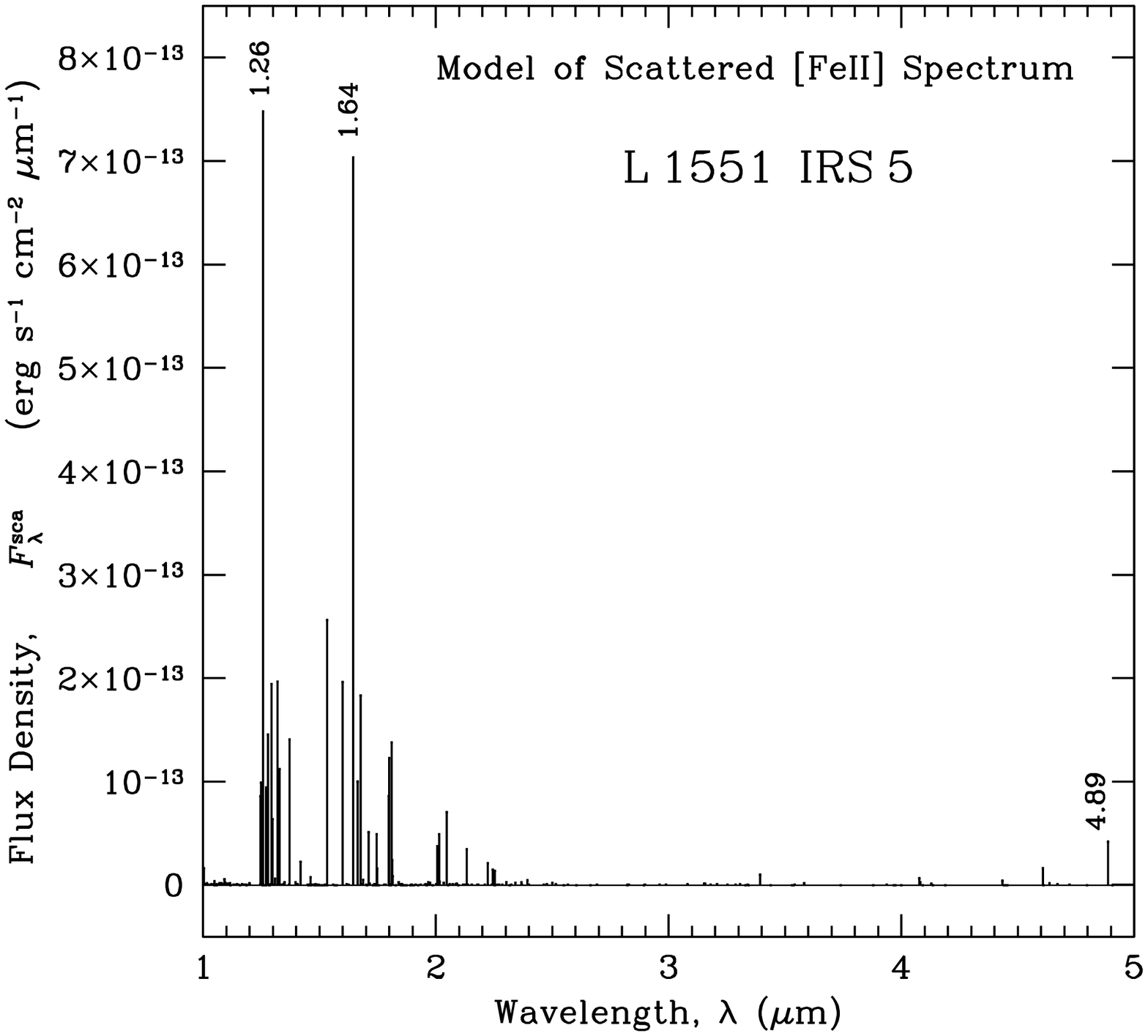}
\end{center}
\caption[]{1 to 5\,$\mu $m spectrum of the source model for L\,1551 IRS\,5 in [Fe\,II] line emission,
based on ISO-SWS observations (see White et al. 2000). Shown is an estimation of the scattered 
light from the central $1^{\prime \prime}$. The spectral resolution is $2 \times 10^4$, corresponding 
to 10 resolution elements of the assumed line width of 150\,km s$^{-1}$.
The slit should be positioned in such a way that the dust scattered [Fe\,II] lines are not 
confused with the emission from the binary jet (Fridlund \& Liseau 1998; Itoh et al. 2000; Pyo et al. 2002).
The strongest lines, in the photometric $J$- and $H$-bands ($1.26\,\mu$m and $1.64\,\mu$m, respectively), 
would be readily accessible from the ground (CRIRES)}
\label{feii_sca}
\end{figure}

Infrared observations of outflows have traditionally focussed on $K$-band ro-vibrational lines of
H$_2$. More recently, also other constituents, notably forbidden lines of ionised iron, have been 
used to trace the associated shocks. These lines are sensitive to temperatures largely in excess
of $10^3$\,K. A nice account for the exploited techniques can be found in the recent
article by Giannini et al. (2004). In addition, Nisini et al. (2004) report on recent IR-H\,I results. 
Needless to say that such studies would largely benefit from increased spectral resolving power,
permitting the observation of the line shapes.

Below, a few observational scenarios are illuminated which have bearing on the open problems 
of the previous sections. The focus will be on the upcoming ESO facilities CRIRES and VISIR, but the 
synergetic and complementing effects of future instruments will also be briefly mentioned.

\subsection{Outflow Origin and their Driving Sources}

High velocity outflows/jets are potential sources of forbidden iron line emission from gas behind $J$-shocks.
There are several infrared lines, which originate from a common upper level, e.g.
$\lambda 1.26$ and $\lambda 1.64$ (CRIRES) on one hand, and $\lambda 4.89$ and $\lambda 17.9$ (CRIRES and VISIR)
on the other (Fig.\,\ref{feii_levels}). The emission is optically thin, hence the intensity ratios of these lines
are constant (for H-neutral gas, $R_{\lambda 1}=I(1.26)/I(1.64)=0.79$ and $R_{\lambda 2}=I(17.9)/I(4.89)=31.7$). 
This circumstance permits the direct assessment of the dust extinction and its spatial variation, in particular 
for highly extinguished sources and/or double/multiple objects (e.g., $\Delta A_{\lambda 1}=0.15$\,mag and 
$\Delta A_{\lambda 2}=-0.08$\,mag, respectively; see White et al. 2000 for the assumed grain composition and 
size distributions).

Toward driving sources of outflows, dust optical depths can become so high that scattering in the 
IR cannot be ignored (above, the extinction estimations represent of course the sum of both absorption and scattering). 
White et al. (2000) have argued that such is the case toward the archetypical outflow source L\,1551\ IRS\,5.
In Fig.\,\ref{feii_sca}, the modelled spectrum of scattered [Fe\,II] from the very central regions, subtending
only about 0$\stackrel {\prime \prime}{_{\bf \cdot}}$2, is shown. The precise nature of the model is critically 
dependent on the correct description of the line profile, requiring high spectral resolution, 
in addition to high sensitivity and angular resolution. Spectropolarimetry would be the desired option, but also
spatially resolved spectroscopy of the emission, scattered up through the excavated polar regions of the
circumstellar disk, could be used for the measurement of the rotation pattern of jets close to their base, i.e. in 
the immedite vicinity of their driving sources. 

\begin{figure}
\begin{center}
\includegraphics[width=0.8\textwidth]{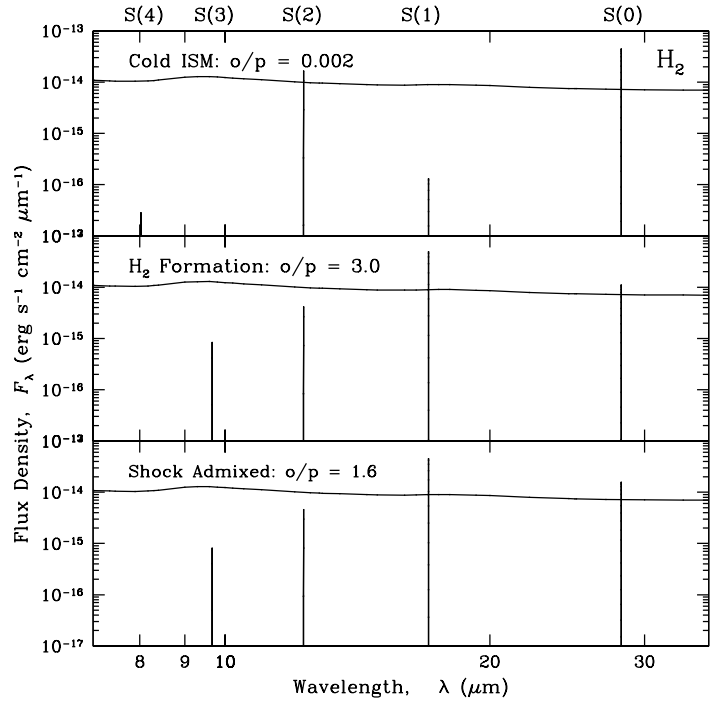}
\end{center}
\caption[]{Observations of rotational H$_2$ lines with differing nuclear spin states potentially 
provide clues to the history of the gas.
The lines are identified above the top panel. The absolute flux scale along
the $y$-axis pertains to a particular model of a parcel of gas at temperatures between
30 and 250\,K and the mass of which is $1\,M_{\rm Jupiter} (D/120\,{\rm pc)^2}$ (the emission is optically thin and
thus easily re-scalable).  
The relative effects of various assumptions about the ortho-to-para ratio, o/p, are readily seen.
In these panels, `Cold ISM' refers to thermally relaxed gas at initially about 20\,K, 
`H$_2$ Formation' would correspond to gas at high temperatures ($> 1\,000$\,K) and
`Shock Admixed' to a mixture of cold and shock-heated gas (this o/p value could also
correspond to gas thermalised at about 100\,K). To guide the eye, an arbitrarily off-set 
extinction curve is included for comparison. Intrinsic line widths are
$3\,{\rm km\,s}^{-1}$ and the spectral resolution is $R=2.5 \times 10^4$
for $\lambda \le 14\,\mu$m and $R=1.25 \times 10^4$ for longer wavelengths (VISIR)}
\label{h2_profile}
\end{figure}

\subsection{Temporal Evolution of the Gas}

The pure-rotational H$_2$ transitions, which, in principle, are admitted by VISIR are 
the lines S(1)\,17.0\,$\mu$m, S(2)\,12.3\,$\mu$m, S(3)\,9.7\,$\mu$m and S(4)\,8.0\,$\mu$m, where
even-$J$ values (in parentheses) refer to para-states and odd-$J$ to ortho-states. 
Being situated close to a silicate feature, the S(3) line could become suppressed by 
circum-/interstellar extinction. However, this line falls onto the wing of deep telluric ozone 
absorption and is hardly accessible from the ground. Generally, the high thermal background 
due to the atmosphere and the warm optics makes high spectral resolution a requirement.

In Fig.\,\ref{h2_profile}, the relative effects of different ortho-to-para ratios of H$_2$,
on an otherwise identical model, are visualised. Obviously, pristine gas would hardly show up
in the ortho lines, whereas these could dominate the emission from more evolved environments,
as, for instance, in the S(1) line. The correct assessment of the H$_2$ ortho-to-para ratio is 
thus instrumental for the understanding of the evolutionary state of the gas.

\subsection{Hot Water from Outflows and Disks}

As already noted, outflows are intimately related to circumstellar disks. Of considerable interest
is the time evolution of the volatile material in outflows and disks and, in particular, that of water. 
Recent, successful ground based observations of fluorescent H$_2$O from comets in the near infrared 
(Dello Russo et al. 2004 and references therein) could also be breaking the ground for such observations 
of disks and outflows, being situated in a high radiation density environment. Such could be encountered
in regions of e.g. high-mass star formation. The observed lines 
originate from highly excited levels, which are practically not populated in the telluric atmosphere.

The high spectral resolution offered by CRIRES could permit the observation of the H$_2$O steam lines 
at about 2.9\,$\mu$m (radiatively excited 200 and 101 bands, decaying to 100 and 001) 
from irradiated outflow gas or externally heated disks (e.g. the proplyds seen 
in silhouette against the bright Orion Nebula). For the latter, preliminary estimates result in emission 
line fluxes of the order of a few times $10^{-15}$\,erg s$^{-1}$ cm$^{-2}$ for a number of the brightest
proplyds, which should be readily observable with CRIRES, as the lines can be assumed to be narrow (a few km s$^{-1}$).  

\section{CRIRES and VISIR in Perspective}

The new ESO facilities CRIRES and VISIR will of course not be operating in complete isolation, 
but provide also valuable complements for current and future instrument developments. In the time 
frame of 2007/08, the spaceborne Herschel Observatory will come into operation. Its 3.5\,m telescope 
will limit the angular resolution at far infrared and submm wavelengths to a few to some tens of arcsec, and
will thus not be able to compete with the VLT in this respect. On the other hand, its heterodyne 
instrument HIFI will offer very high spectral resolution, in excess of $10^6$. On a somewhat longer 
time scale, the large mm/submm array ALMA will offer both high angular (0$\stackrel 
{\prime \prime}{_{\bf \cdot}}$1) and spectral resolution 
($\gg 10^5$). At shorter wavelengths, the 6\,m James Webb Space Telescope (JWST) will provide 
diffraction limited IR observations (at $2\,\mu$m), in addition to very high sensitivity, especially
in the thermal infrared, thanks to the low-noise environment at the Sun-Earth $L2$. However, 
the direct, spatial resolving of the outflow acceleration regions, at the level of a few milliarcsec 
and with JWST sensitivity, will probably need to await the advent of the InfaRed Space Interferometer Darwin, 
sometime in the middle of the next decade.

\section{Conclusions}

\begin{itemize}
\item[$\bullet$] The ubiquitous phenomenon of outflows in star forming regions is discussed. 
Our knowledge appears still largely incomplete, in particular what regards fundamental issues, such as 
outflow origin and collimation.
\item[$\bullet$] High resolution IR spectral mapping observations can potentially put models of outflow
origin to test by measuring the rate of jet rotation. In highly extinguished regions, scattered lines 
of [Fe\,II] are likely suitable to accomplish this.
\item[$\bullet$] The observation from the ground of the rotational lines of H$_2$ in the thermal infrared  
requires high spectral and spatial resolution. The relative line strengths of ortho-H$_2$ and para-H$_2$ 
can be used to study the history of shocked/non-shocked interstellar gas in outflows and disks.
\item[$\bullet$] High resolution IR observations of lines of hot H$_2$O open up the possibility to
study the H$_2$O evolution in disks and outflows from the ground. 
\end{itemize}

\noindent
{\it Acknowledgements.} I enjoyed interesting discussions with M.\,Barsony, A.\,Caratti o Garatti, B.\,Nisini 
and G.\,Olofsson. I would also like to take this opportunity 
to thank the organisers for their kind hospitality during this stimulating workshop.

\end{document}